\documentclass[%
reprint,
superscriptaddress,
 amsmath,amssymb,
 aps,
]{revtex4-1}

\usepackage{graphicx}% Include figure files
\usepackage{dcolumn}% Align table columns on decimal point
\usepackage{bm}% bold math
\usepackage[flushleft]{threeparttable} % table footnote
\usepackage{multirow}

\usepackage{dsfont}
\usepackage{color,psfrag}
\usepackage{mathtools,amscd}

\usepackage{soul}
\usepackage{xcolor} % colored tables

\usepackage{ifpdf}
    \ifpdf
\usepackage{tikz}
\usetikzlibrary{arrows,chains,matrix,positioning,scopes, fit, calc}
\usetikzlibrary{cd}
\usepackage{chemfig}
\fi
\usepackage{hyperref}
\usepackage{float}

\hypersetup{
	colorlinks,
	linkcolor=blue,
	citecolor=blue, 
	filecolor=black,
	urlcolor=blue}

\newcommand{\mat}[1]{\mathbf{#1}} % matrices in bold

\begin{document}

\preprint{APS/123-QED}

%%%%%%%%%%%%%%%%%%%%%%%%%%%%%%%%%%%%%%%%%%%%%%%%%%%%%%%%%%%%%%%%%%%
%%%%%%%%%% =========================== T I T L E ============================%%%%%%%%%%%%%%%%
%%%%%%%%%%%%%%%%%%%%%%%%%%%%%%%%%%%%%%%%%%%%%%%%%%%%%%%%%%%%%%%%%%%

\title{Molecular Force Fields with Gradient-Domain Machine Learning (GDML): Comparison and Synergies with Classical Force Fields}% Force line breaks with \\
\author{Huziel E. Sauceda}
\email{sauceda@tu-berlin.de}
 \affiliation{%
 Department of Physics and Materials Science, University of Luxembourg, L-1511 Luxembourg, Luxembourg
}%
\affiliation{%
 Machine Learning Group, Technische Universit\"at Berlin, 10587 Berlin, Germany
}%
\affiliation{%
 BASLEARN, BASF-TU joint Lab, Technische Universit\"at Berlin, 10587 Berlin, Germany
}%
\author{Michael Gastegger}
 \affiliation{%
 Machine Learning Group, Technische Universit\"at Berlin, 10587 Berlin, Germany
}%
\affiliation{%
	DFG Cluster of Excellence ``Unifying Systems in Catalysis'' (UniSysCat), Technische Universit\"at Berlin, 10623 Berlin, Germany
}%
\affiliation{%
	BASLEARN, BASF-TU joint Lab, Technische Universit\"at Berlin, 10587 Berlin, Germany
}
\author{Stefan Chmiela}
 \affiliation{%
 Machine Learning Group, Technische Universit\"at Berlin, 10587 Berlin, Germany
}%
\author{Klaus-Robert M\"uller}%
 \email{klaus-robert.mueller@tu-berlin.de}
\affiliation{%
 Machine Learning Group, Technische Universit\"at Berlin, 10587 Berlin, Germany
}
\affiliation{%
Department of Brain and Cognitive Engineering, Korea University, Anam-dong, Seongbuk-gu, Seoul 136-713, Korea
}%
\affiliation{%
Max Planck Institute for Informatics, Stuhlsatzenhausweg, 66123 Saarbr\"ucken, Germany}
\affiliation{%
Google Research, Brain team, Berlin, Germany
}
\author{Alexandre Tkatchenko}
 \email{alexandre.tkatchenko@uni.lu}
 \affiliation{%
 Department of Physics and Materials Science, University of Luxembourg, L-1511 Luxembourg, Luxembourg
}%
\date{\today}% It is always \today, today,
%\myworries{[benzene, uracil, naphthalene, aspirin, salicylic acid, malonaldehyde, ethanol, toluene]}
\date{\today}% It is always \today, today,
%\myworries{[benzene, uracil, naphthalene, aspirin, salicylic acid, malonaldehyde, ethanol, toluene]}
\begin{abstract}
Modern machine learning force fields (ML-FF) are able to yield energy and force predictions at the accuracy of high-level \emph{ab initio} methods, but at a much lower computational cost.
On the other hand, classical molecular mechanics force fields (MM-FF) employ fixed functional forms and tend to be less accurate, but considerably faster and transferable between molecules of the same class. In this work, we investigate how both approaches can complement each other. We contrast the ability of ML-FF for reconstructing dynamic and thermodynamic observables to MM-FFs in order to gain a qualitative understanding of the differences between the two approaches. This analysis enables us to modify the generalized AMBER force field (GAFF) by reparametrizing short-range and bonded interactions with more expressive terms to make them more accurate, without sacrificing the key properties that make MM-FFs so successful.
\end{abstract}
\pacs{Valid PACS appear here}% PACS, the Physics and Astronomy
\maketitle

%\tableofcontents

%%%%%%%%%%%%%%%%%%%%%%%%%%%%%%%%%%%%%%%%%%%%%%%%%%%%%%%%%%%%%%%%%%%
%%%%%%%%%% ========================== INTRODUCTION ========================== %%%%%%%%%%%%%
%%%%%%%%%%%%%%%%%%%%%%%%%%%%%%%%%%%%%%%%%%%%%%%%%%%%%%%%%%%%%%%%%%%

\section{Introduction}
% General intro
Computational studies of materials and molecular systems at the atomic level of detail constitute a central tool in contemporary physics, biology, materials science, and chemistry research.
Standard electronic-structure theories (e.g. density-functional theory (DFT) or wavefunction-based methods) are ultimately limited in the size of systems and time scales they can treat, due to the high computational cost associated with modeling the electronic interactions.
Conventional molecular mechanics force fields (MM-FFs) overcome these limitations by introducing a range of computationally efficient approximations, circumventing the need to solve the electronic problem explicitly~\cite{BIXON1967,AMBER1981,CHARMM1983,MMFF94,GROMOS2005,TIP4P1983,TIP5P2000,gaff2004}.
This has made possible to simulate biologically relevant systems (such as proteins), gaining insights into their innermost workings at an atomistic level, culminating a Nobel prize in chemistry awarded to Karplus, Levitt and Warshel.
However, the high computational efficiency of MM-FFs comes at the cost of accuracy compared to electronic-structure methods. Moreover, the basic form of the physical approximations employed restricts their applicability to certain classes of problems, e.g. most of them are unable to model chemical reactions involving bond breaking/formation or do not include crucial intramolecular interactions~\cite{sGDMLappl1} or many-body effects~\cite{Stoehr2019}.

% Intro ML
Recently, modern machine learning methods (ML) have emerged as an elegant solution to this dilemma \cite{QML-Book,von2020exploring,noe2020machinemolsim}.
Due to the high functional flexibility of ML methods they can be used to construct force fields (FFs) with the accuracy of electronic-structure calculations, while retaining high computational efficiency albeit still behind conventional classical force fields.
A plethora of ML based FFs has been reported within the past few years, demonstrating the general utility of this approach. 
In particular, a great amount of work has been done in FFs based on neural networks (NN)~\cite{Behler2007,Behler2007b,Behler2011,Behler2011a,Behler2012,Behler2016,Gastegger2017,dtnn,SchNetNIPS2017,SchNet2018,unke2019physnet} and kernel-based models~\cite{Bartok2013,Bartok2015_GAP,Ramprasad2015,DeVita2015,Shapeev2017,Glielmo2017,gdml,sgdml,sGDMLappl1,wang2020ensemble}.
In order to understand better how these models work and to improve their performance, other ML studies have been done on molecular representations~\cite{Rupp2012,Bartok2010,Hansen2013,Hansen2015,Rupp2015,Ceriotti2016,artrith2017efficient,Ceriotti2017,yao2017many,faber2017prediction,eickenberg2018solid,glielmo2018efficient,Grisafi2018,tang2018atomistic,pronobis2018many}, data sampling~\cite{dral2017structure,noe2018,noe2018b,Subramanian2019}, explanation methods~\cite{dtnn,meila2018}, and inference of molecular properties~\cite{Bartok2013,Montavon2013a,Ramprasad2015,Brockherde2017,huan2017universal,Tristan2018,lubbers2018hierarchical,kanamori2018exploring,hy2018predicting,Smith2017,Clementi2018,winter2019,gdml,sgdml}, as well as software development~\cite{sGDMLsoftware2019,yao2018tensormol,schnetpack2018}.
Yet, although it is generally assumed that ML-FFs have much better accuracy than MM-FFs, no in-depth comparisons (e.g. potential-energy surfaces and molecular dynamics predictions) have been performed up to date.
Furthermore, such comparison raises the question of how the ML-FFs and MM-FFs could be combined to complement each other.  

The goal of the present work is to perform a detailed investigation of the differences between both approaches based on a set of small molecules exhibiting different quantum-mechanical phenomena.
Based on these results, different alternatives are explored for improving the data generation process and their applicability for expediting the force-field learning procedure. 
Furthermore, improvement of the accuracy for MM-FFs is studied by reparameterising them based on more accurate reference data and test their limits and functional form flexibility.
For this task, we use the sGDML framework \cite{gdml,sgdml} as ML-FF of choice, as it is able to efficiently reconstruct the potential-energy surfaces (PES) of medium sized molecules.
The investigated systems are the molecules ethanol, the keto form of malondialdehyde (keto-MDA) as well as acetylsalicylic acid (Aspirin). 
In the context of these systems, we study the performance of MM-FFs and sGDML derived FFs based on the overall reliability of the generated PESs, as well as effects arising from chemical phenomena such as interactions between molecular orbitals.
Although we restrict ourselves to the sGDML approach, it can nevertheless be expected that the results found here are equally valid for ML-FFs in general.

%%%%%%%%%%%%%%%%%%%%%%%%%%%%%%%%%%%%%%%%%%%%%%%%%
%============= THEORY ==========================%
%%%%%%%%%%%%%%%%%%%%%%%%%%%%%%%%%%%%%%%%%%%%%%%%%

% ==================== F I G U R E ===================
\begin{figure*}[ht]
%\centering
\includegraphics[width=\textwidth]{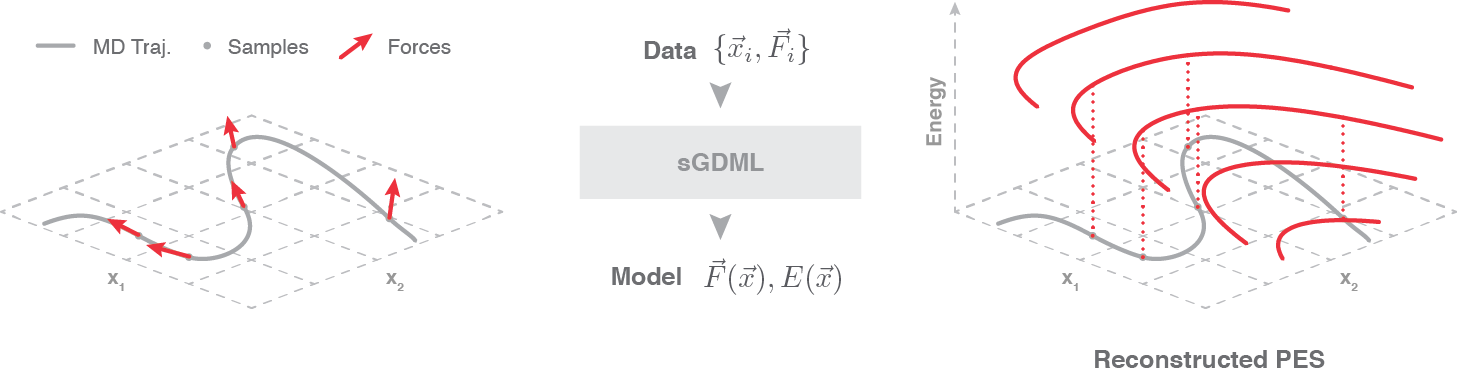}

\caption{Construction of the sGDML model. (Left) From a molecular dynamics trajectory (MD Traj.) a random set of data points, \{\textbf{x}$_i$, \textbf{F}$_i$\}$_{i=1}^M$, is collected (\textbf{x}: gray dots, \textbf{F}: red arrows). (Middle) The \{\textbf{x}$_i$, \textbf{F}$_i$\}$_{i=1}^M$ data is then the input for the sGDML software, which the provides the reconstructed force field and its potential-energy surface (PES) (Right).}
\label{fig:FigModel}
\end{figure*}
% =====================================================

\section{Theory} \label{Theory}
Learning theory provides a framework to create a diverse collection of universal approximators capable of to reconstruct any function. 
In practice, reconstructing PESs from highly accurate reference calculations such as coupled cluster theory (CC) is a challenging task given that the amount of available data is limited by computational resources that such level of theory requires.
Consequently, mathematically constrain the solution space of a ML approximator by enforcing universal physical laws it is highly advantageous.
In this manner, a data-efficient model capable of delivering physically meaningful predictions can be obtained.
%
% TODO: This is if you want an "accurate" ML thingy
Some of the basic desirable properties of high accuracy ML-FFs, from the view of physics and chemistry, are:

\textit{A global description of the system}. The nature of the quantum interactions resulting from the Schr{\"o}dinger equation (within the Born-Oppenheimer approximation) $\mathcal{H}\Psi=V_{BO}\Psi$ and from the evaluation of the Hellmann-Feynman forces $-\textbf{F}=\langle \Psi^*|\partial \mathcal{H} /\partial \textbf{x}|\Psi \rangle$ is inherently many-body. 
In practice, such global description of the system has to be enforced by avoiding the non-unique partitioning of the total energy $V_{BO}$ in atomic contributions.

\textit{Time invariance of the Hamiltonian}. This constraint requires a time-invariant Hamiltonian $\mathcal{H}= \mathcal{T}+\hat{f}_E$, hence energy conservation must be enforced in the ML model, $\mat{\hat{f}}_{\mat{F}} = - \nabla \hat{f}_E$, yielding conservative forces. 

\textit{Permutational invariance of identical atoms}. In quantum mechanics, the square of the wave function is invariant to permuting two identical atoms in a molecule, a property that is carried over to the energy of the system: $V_{BO}(\ldots,{\color[rgb]{0.000111,0.001760,0.998218}\vec{x}_i},\ldots,{\color[rgb]{0.986246,0.007121,0.027434}\vec{x}_j},\ldots)=V_{BO}(\ldots,{\color[rgb]{0.986246,0.007121,0.027434}\vec{x}_j},\ldots,{\color[rgb]{0.000111,0.001760,0.998218}\vec{x}_i},\ldots)$. 
Fulfilling this spatial symmetry by the ML models often represents a big challenge, as well as in MM-FFs since a key ingredient of ML models is to invariantly compare objects, here molecules. 
Thus, if no invariance can be guaranteed then permutations typically give rise to noisy comparisons hampering consistent learning.
The above mentioned physical properties constitute a set of mathematical constraints that narrows down the space of solution of the general ML approximator, thereby increasing the data efficiency and accuracy in the reconstruction of the original data generator.
However, all these considerations, and the desire for a global model in particular, place limitations to the computational performance of such an approach.
%As could be expected, this approach faces some computational limitations performance when compared to the lighting fast MM-FFs calculations.
%

A completely different strategy is pursued by molecular mechanical force fields, whose design follows chemical intuition and known physical approximations to intra- and intermolecular interactions.
In this case, the globality of the model is replaced by a more scalable partitioning of the interactions in terms of interatomic pair distances, angles and dihedral angles in local domains.
The 3-body (angles) and 4-body (dihedral angles) terms are the ones responsible for the permutational invariance violation in MM-FFs, a problem that becomes less and less relevant as the size of the system increases from 100s to 1000s or even millions of atoms.
On the other hand, these force fields are energy conserving by construction, since the molecular forces are derived as the derivative of a potential in the form of the interaction energy.
This strategy has been proven to be highly successful over the years in many domains of molecular modeling~\cite{AMBER1981}.

%TODO: This is strange
Before comparing these two contrasting methodologies, we briefly review both frameworks, which will be important for the interpretation and further improvement of both approaches.

\subsection{\texorpdfstring{\MakeLowercase{s}}~GDML model} \label{sGDMLmodel}

The requirements mentioned above served as a basis for the \textit{\textbf{s}}ymmetric \textit{\textbf{G}}radient \textit{\textbf{D}}omain \textit{\textbf{M}}achine \textit{\textbf{L}}earning (sGDML) FF~\cite{sgdml,sgdml_bookTheory}. 
This framework consisting in the use of a permutationaly-symmetrized Hessian of the covariance function, which then satisfy energy conservation and atomic indistinguishability by construction. 
The global nature of the interactions is retained by using a global molecular descriptor.

In practice, the sGDML model consisting in the gradient of a kernel ridge regression (KRR) estimator is trained only on forces $\mathbf{F}$, and the use of the Hessian matrix of the kernel $\kappa$ as a covariance structure, naturally gives an analytically integrable force estimator $\mat{\hat{f}_F}$, $\hat{f}_E=\int\mat{\hat{f}_F}\cdot d\mat{x}+c$~\cite{gdml,sgdml}. 
Then, the training of the model consists in solving the normal equation of the KRR estimator in the gradient domain for the parameter-vectors $\vec{\alpha}$~\cite{gdml,sgdml}:
\begin{equation}
\left(\mathbf{K}_{\text{Hess}(\kappa)} + \lambda \mathbb{I}\right) \vec{\alpha}= \nabla V_{BO} = -\mathbf{F} \text{,}
\end{equation}
\noindent here $\mathbf{K}_{\text{Hess}(\kappa)}$ is the kernel matrix and $\mathbb{I}$ is the identity matrix weighted by the regularization parameter $\lambda$. 
An extensive description of the sGDML framework can be found in Refs.~\citenum{sgdml}, ~\citenum{sgdml_bookTheory} and \citenum{sGDMLsoftware2019}.

The atomic indistinguishability is imposed by adding permutational symmetry (i.e. rigid space group and fluxional symmetries) constraints on $K_{\text{Hess}(\kappa)}$~\cite{sgdml}. 
Extracting those symmetries from the molecular structure requires physical and chemical intuition about the system under study, e.g. knowing if a molecular rotor could jump the energetic rotational barriers at a given temperature, which is impractical in general. 
This problem is circumvented by implementing a data-driven multi-partite matching procedure that automates the discovery of permutation matrices $\mat{P}(\tau)$ corresponding to the index permutation $\tau$ embedded in the training data, e.g. molecular dynamics simulations trajectories, followed by a synchronization procedure. 
The permutation group recovered by the multi-partite matching procedure, also known as the \textit{Longuet-Higgins group}~\cite{Longuet-Higgins1963}, has the particular advantage that it limits the symmetry recovery to only energetically feasible permutational configurations. 
This avoids considering unfeasible permutations such as the permutation of two random atoms of the same species which would not contribute any valuable information to the symmetrized kernel. 
This translates into a strong reduction of computational efforts in evaluating the model.

The final form of the FF estimator trained on $M$ reference geometries, for a molecule with $N$ atoms and a Longuet-Higgins group containing $S$ symmetry transformations, takes the form
\begin{equation}
\mat{\hat{f}_F}(\vec{x}) = \sum^M_{i} \sum^{S}_{q} \{(\mat{P}_q \vec{\alpha}_{i})  \cdot \nabla\}\nabla \kappa(\vec{x},\mat{P}_{q}\vec{x}_i)\,.
\label{eq:force_model}
\end{equation}
The corresponding energy $\hat{f}_E$ predictor describing the underlying PES is obtained by integrating $\mat{\hat{f}_F}$ with respect to the Cartesian coordinates, 

\begin{equation}
-\hat{f}_{\text{E}}(\vec{x}) = \int \mat{\hat{f}_F} \cdot d\mat{x} = \sum^M_{i} \sum^{S}_{q} \{(\mat{P}_q \vec{\alpha}_{i}) \cdot \nabla\}\kappa(\vec{x},\mat{P}_{q}\vec{x}_i)\,.
\label{eq:force_model}
\end{equation}

Figure~\ref{fig:FigModel} gives a general perspective of the sGDML model's training process, from sampling the molecular dynamics trajectory and the permutational symmetries extracting and reconstruction process of the the underlying PES. 

Recently, we have systematically demonstrated that sGDML models trained on only few 100s of reference structures are able to reconstruct molecular PESs with a mean absolute error (MAE) of less that 0.06 kcal~$\text{mol}^{-1}$ for small molecules with up to 15 atoms and less than 0.16 kcal~$\text{mol}^{-1}$ for molecules as complex as aspirin, paracetamol, and azobenzene~\cite{sGDMLappl1,sgdml,sGDMLsoftware2019,sgdml_bookAppl,NQE_vdW_2020}. These results will be of particular importance when directly comparing with less flexible MM-FFs in section~\ref{Reparametrization}.

\subsection{Molecular Mechanics Force Fields} \label{MMFFs}

Molecular mechanics force fields are designed with the goal to render simulations possible which are intractable with conventional electronic-structure methods, either due to system size or simulation time scales.
This is achieved by substituting quantum-mechanical interactions with simpler, computationally efficient physical approximations.

In order to construct these approximations, most common force fields explicitly consider the bonds between the atoms in a molecule.
Hence, based on the seminal work by Bixon and Lifson~\cite{BIXON1967}, the total energy predicted by a FF can be separated into a sum of bonded and non-bonded interactions:

\begin{equation}
E_\mathrm{FF} = E_\mathrm{bonded} + E_\mathrm{non-bonded}
\end{equation}

% ======= bonded ==========

Bonded interactions describe the energy contributions due to the interactions of atoms with their closest neighbors and are typically expressed in terms of internal coordinates (bonds, angles and dihedral angles).
Bond energies are modelled after experimentally observed dissociation curves for diatomic molecules via truncated Taylor series. For the sake of computational efficiency, these series are truncated after the quadratic term, leading to the harmonic expression:

\begin{equation}
E_{\mathrm{bond}}(r_{ij}) = \frac{1}{2} k_{ij} \left( r_{ij} - r_{ij}^{(\mathrm{eq})} \right)^2,\label{eq:bondharm} 
\end{equation}

\noindent where $r_{ij}$ is the distance between atoms $i$ and $j$, $r_{ij}^{(\mathrm{eq})}$ is the equilibrium bond length and $k_{ij}$ is the force constant modulating the strength of the bond. $k_{ij}$ and $r_{ij}^{(\mathrm{eq})}$ differ depending on the chemical elements involved in the bonds, as well as their chemical environments (e.g. hybridization state) and are determined based on experimental observations or electronic-structure calculations.

Angles are modeled in a similar manner:

\begin{equation}
E_{\mathrm{angle}}(\alpha_{ijk}) = \frac{1}{2} k_{ijk} \left( \alpha_{ijk} - \alpha_{ijk}^{(\mathrm{eq})} \right)^2 \label{eq:angharm}
\end{equation}

\noindent Here, $\alpha_{ijk}$ is the angle between the bonded atoms $i$, $j$ and $k$, while $\alpha_{ijk}^{(\mathrm{eq})}$ and $k_{ijk}$ are the angle at equilibrium and the associated restitution force constant, respectively.
Interactions arising from dihedral angles are modeled via Fourier series in order to account for the correct periodicity of the associated PES:

\begin{align}
E_{\mathrm{dihedral}}(\theta_{ijkl}) =& \frac{1}{2} \sum_{\gamma} V_{ijkl}^{(\gamma)} \times \\
 &\left[ 1 + (-1)^{\gamma+1} \cos(\gamma \theta_{ijkl} + \phi_{ijkl}^{(\gamma)} )\right],\label{eq:dihed}
\end{align}

\noindent where $\theta_{ijkl}$ is the dihedral angle between atoms $i$, $j$, $k$ and $l$. $\phi_{ijkl}^{(\gamma)}$ is a phase shift and $V_{ijkl}^{(\gamma)}$ is an amplitude controlling the interaction strength.
The total bonded energy is then obtained as sum over all of these terms arising from to the predetermined connectivity of the investigated system:

\begin{align}
E_\mathrm{bonded} &= \sum_{\mathrm{bonds}}                       E_{\mathrm{bond}}(r_{ij}) \\
                  &+ \sum_{\mathrm{angles}}               E_{\mathrm{angle}}(\alpha_{ijk}) \\
                  &+ \sum_{\mathrm{dihedral}}  E_{\mathrm{dihedral}}(\theta_{ijkl})
\end{align}

Non-bonded interactions -- defined as all interactions of atoms separated by more than two neighbors -- are most commonly treated in the form of pair-wise potentials in conventional FFs.
Electrostatic interactions are modeled as a Coulomb potential:

\begin{equation}
E_{\mathrm{Coulomb}}(r_{ij}) = \frac{1}{4\pi\epsilon_0} \frac{q_i q_j}{r_{ij}}.\label{eq:coulomb}
\end{equation}

\noindent $\epsilon_{0}$ is a dielectric constant and $q_i$ and $q_j$ are the partial charges associated with the interacting atoms. These properties are once again free parameters of the FF model and need to be determined based on the identity of the interaction partners. Finally, long range van der Waals type interactions are typically accounted for via a Lennard--Jones potential~\cite{LennardJones1931}, due to its correct decay behavior and computationally efficient evaluation:

\begin{equation}
E_{\mathrm{vdW}}(r_{ij}) = 4 \epsilon_{ij} \left[ \left(\frac{A_{ij} }{r_{ij}}\right)^{12} - \left(\frac{B_{ij} }{r_{ij}}\right)^6 \right].\label{eq:lj} 
\end{equation}

\noindent Here, $\epsilon_{ij}$ controls the depth of the potential well and the parameters $A_{ij}$ and $B_{ij}$ modulate its overall shape.
As was the case with bonded interactions, all non-bonded terms are combined to yield:

\begin{align}
E_\mathrm{non-bonded} &= \sum_{i,j>i} E_{\mathrm{Coulomb}}(r_{ij}) \\
                      &+ \sum_{i,j>i} E_{\mathrm{vdW}}(r_{ij})
\end{align}

\noindent Although the number of bonded energy contributions grows linearly with the number of atoms present in the system, non-bonded interactions grow quadratically due to their pairwise nature.
In order to overcome this $\mathcal{O}(N^2)$ scaling behavior, long range effects are typically truncated after a certain distance, since both -- electrostatic and van der Waals energies -- decay with increasing distances.

At the core of every force field lies a set of parameters tabulated for the above interactions, which are determined by fitting to experimental and theoretical data.
There is a wide range of conventional force fields, differing in the class of target systems and the types of atom environments considered (e.g. AMBER~\cite{AMBER1981}, CHARMM~\cite{CHARMM1983}, GROMOS~\cite{GROMOS2005}).
All of these excel at high computational speeds, making possible to simulate systems containing hundreds of thousands of atoms, e.g. biologically relevant systems such as proteins, DNA strands and cell membranes\cite{wang2001biomolecular}.
However, the price MM-FFs pay for this efficiency is a loss of flexibility and accuracy. For example, harmonic expressions for bond energies (Eq.~\ref{eq:bondharm}) are only valid close to the equilibrium and fail to account for anharmonicities in bonds, as well as situations involving formation and breaking of bonds.

\section{Computational Methods}

All sGDML models in this study were trained on subsets of 1000 points sampled from the respective bulk datasets of reference calculations according to the Boltzmann distribution. Smaller independent validation subsets were used to determine the best hyperparameter-configuration for each model. The remainders of each dataset served as test splits, to estimate the generalization performance of each model. This process is fully automated by the sGDML software package (available at \texttt{www.sgdml.org})
~\cite{sGDMLsoftware2019}.
The molecular datasets used in this study were generated using DFT at the generalized gradient approximation (GGA) level of theory with Perdew-Burke-Ernzerhof (PBE) exchange-correlation functional~\cite{PBE1996} and the the van der Waals interactions were incorporated using the Tkatchenko-Scheffler (TS) method~\cite{TS} as implemented in the FHI-aims package~\cite{FHIaims2009}.
For the keto-MDA, enol-MDA and ethanol molecules we also computed datasets using all-electron CCSD(T) and all-electron CCSD for Aspirin using the Psi4 software~\cite{psi42012,psi42017,psi42018}.

PES scans for the different MM-FFs were performed with the Atomic Simulation Environment (ASE)~\cite{larsen2017atomic}, using the built-in calculator for GAFF and custom calculators for UFF~\cite{UFF}, MMFF94\cite{MMFF94} and Ghchemical\cite{Ghemical}, based on their implementation in openbabel~\cite{o2011open}.
For GAFF computations, the AMBER 18 force field package~\cite{salomon2013overview} was used and the charges of all molecules were derived from RESP charges~\cite{woods2000restrained} at the HF/6-31G* level of theory~\cite{roothaan1951new,petersson1988complete} computed with Gaussian~\cite{g09}.
Molecular dynamics simulations were carried out using ASE. A timestep of 0.5~fs was used to propagate the system and temperatures were kept constant using a Langevin thermostat~\cite{PhysRevE.75.056707}.
sGDML simulations were performed under similar conditions using the sGDML-ASE calculator implemented in the sGDML software package~\cite{sGDMLsoftware2019}.

\section{Results and Discussion}

% ==================== F I G U R E ===================
\begin{figure*}[htp]
	\centering
	\includegraphics[width=0.9\textwidth]{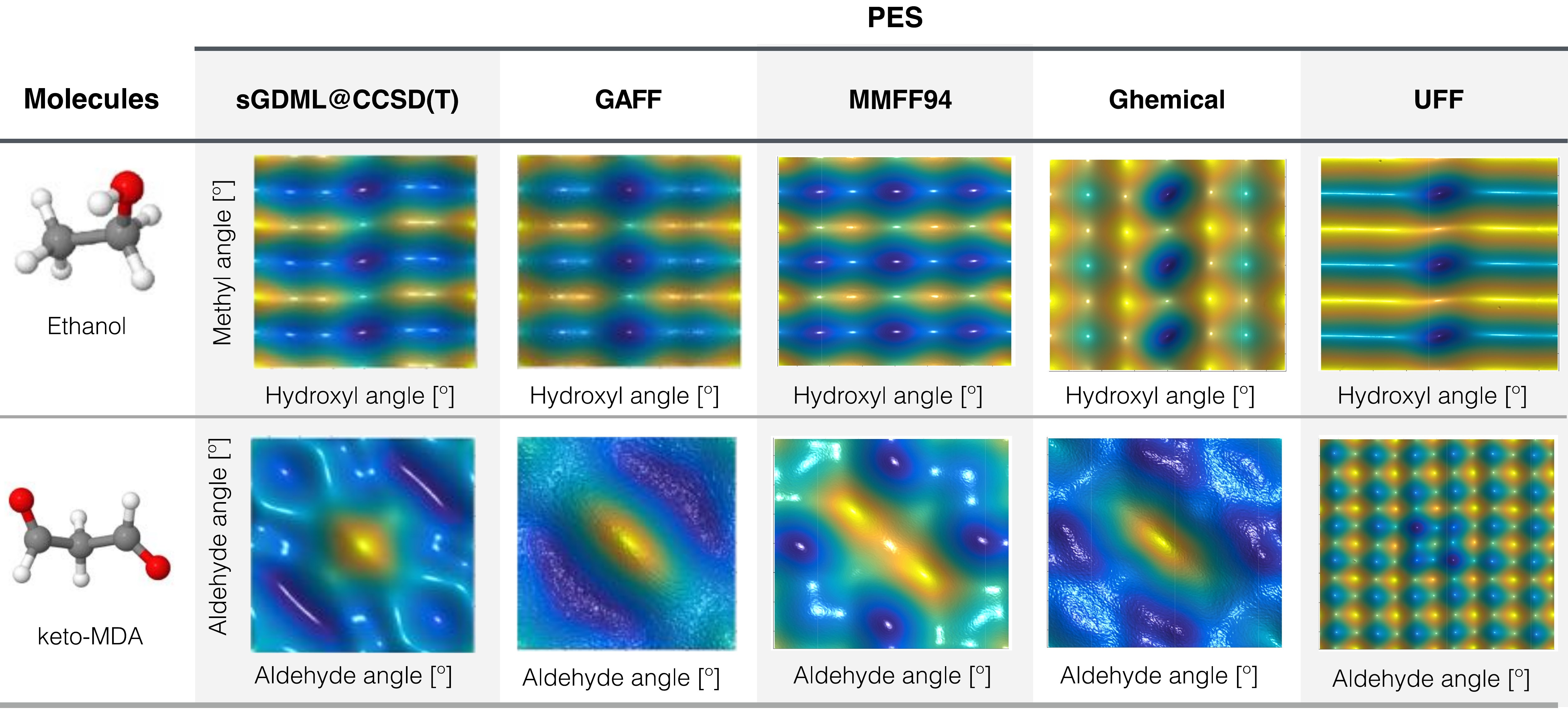}
	\caption{Comparison of PES generated using several MM-FFs for ethanol and keto-MDA compared to the sGDML@CCSD(T) ML-FF. See Supporting Information for more details.}
	\label{fig:FigPESAllFFs}
\end{figure*}
% =====================================================

% ==================== F I G U R E ===================
\begin{figure}[t]
	\centering
	\includegraphics[width=1.0\columnwidth]{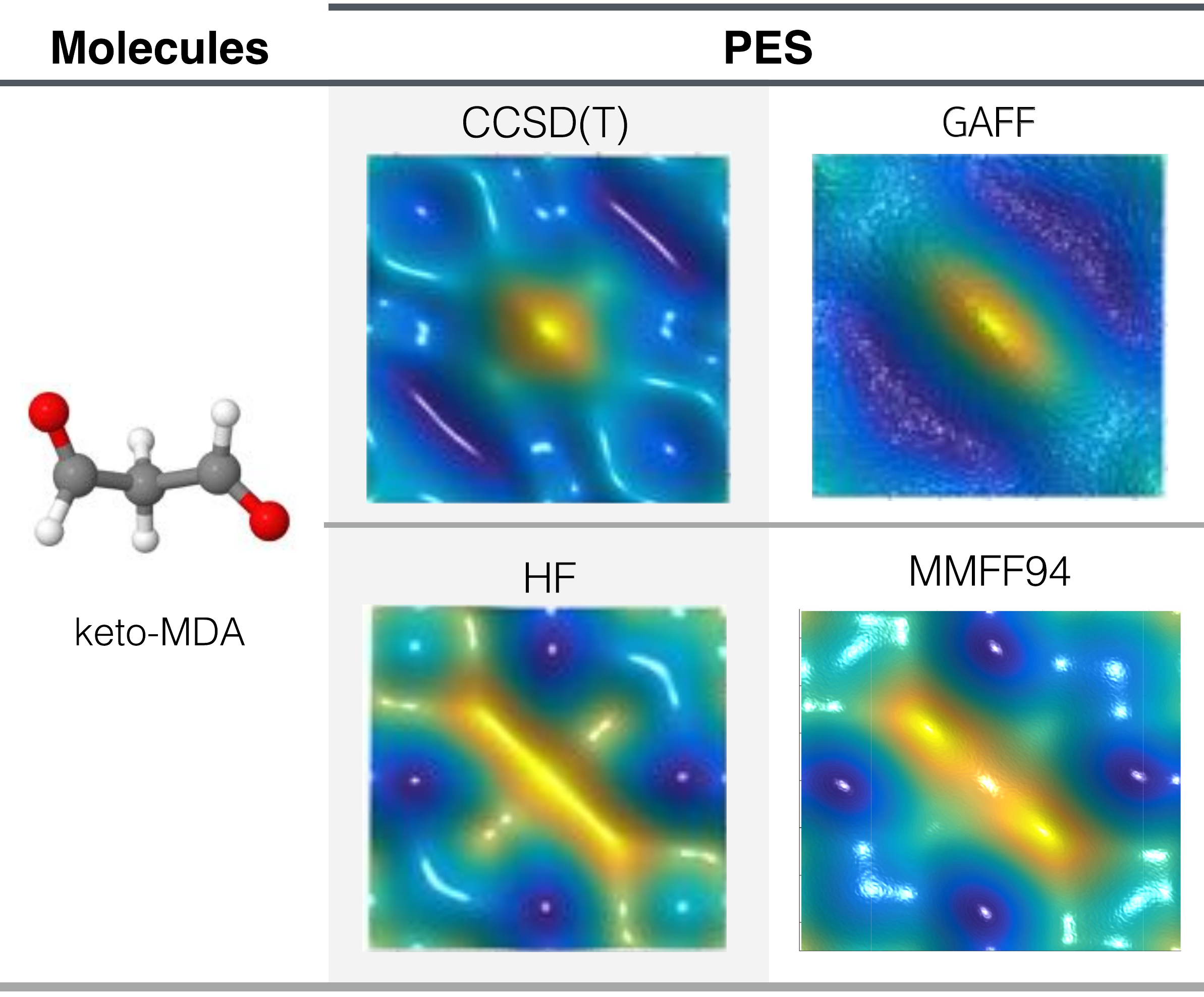}
	\caption{Comparison of keto-MDA's PES generated from sGDML@CCSD(T), GAFF, sGDML@HF, and MMFF94.}
	\label{fig:FigCompHF}
\end{figure}
% =====================================================

\subsection{Comparison of MM-FFs and ML-FFs} \label{PES}

ML-FFs are able to construct efficient and accurate surrogate models of high-level electronic-structure methods.
As a consequence, they can be used to directly compare e.g. CC-level PES with the predictions of conventional MM-FFs, offering insights into the quality and potential shortcomings of these methods at a fundamental level.
This in turn can pave the way towards improved MM-FF approaches.

Here, we demonstrate the utility of such a comparison based on ethanol and keto-malondialdehyde as example systems.
These molecules are very useful to exemplify two ubiquitous phenomena in chemistry and biology, the effects of electron lone-pairs and purely quantum-mechanical effects (e.g. $n\to\sigma^*$ transitions)~\cite{sGDMLappl1,n2pi_Aspirin2011,n-pi2017}.
As a consequence, the PES of both molecules exhibit highly complex features,which even DFT approximations struggle to capture~\cite{sgdml,sGDMLappl1}. 

\begin{figure*}[ht]
	\centering
	\includegraphics[width=0.9\textwidth]{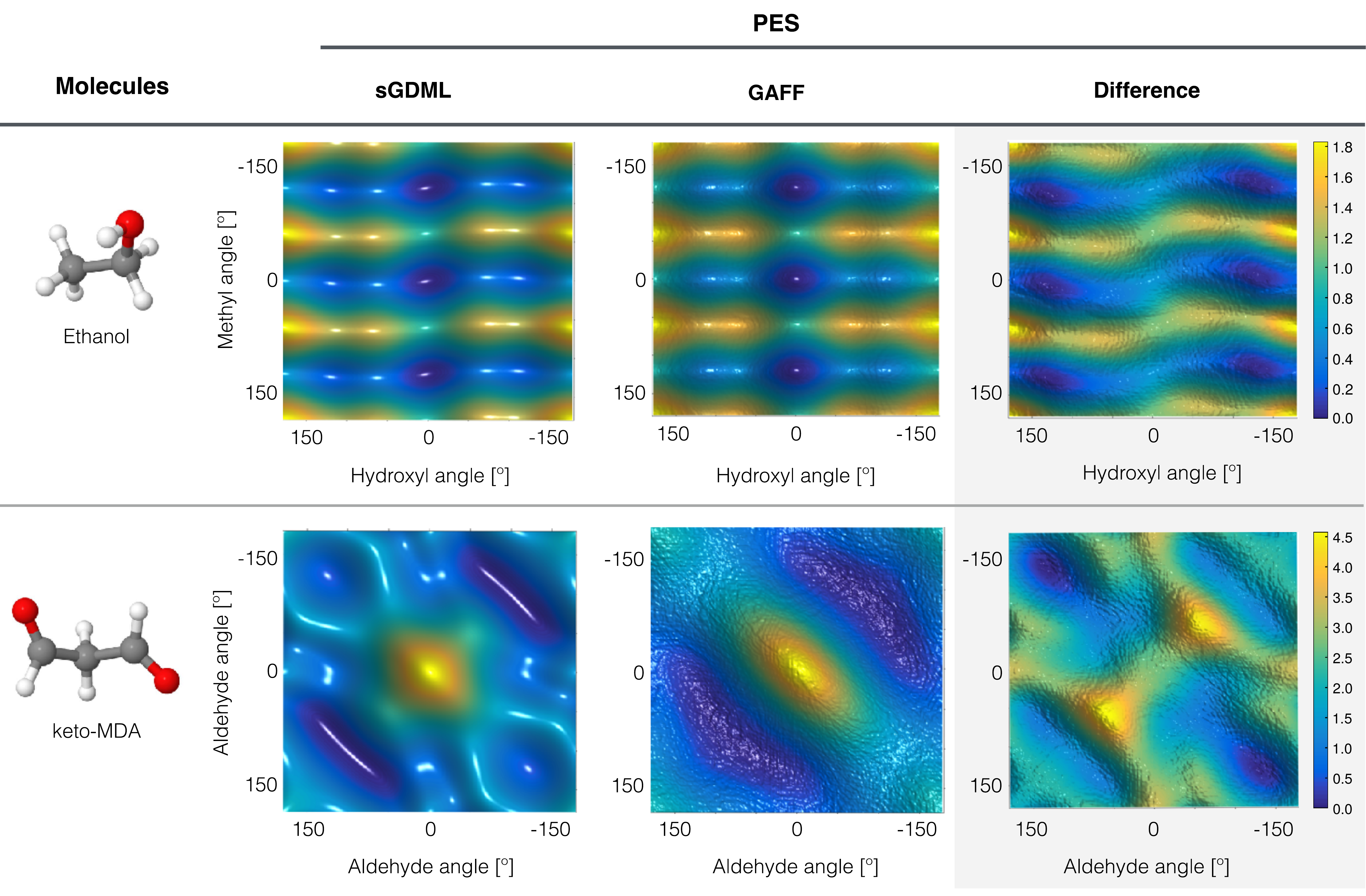}
	\caption{PESs generated from sGDML@CCSD(T)/cc-pVTZ, GAFF for ethanol and keto-MDA, and their respective difference PES$_{\text{sGDML}}$--PES$_{\text{GAFF}}$. Energy differences are given in kcal\,mol$^{-1}$}
	\label{fig:FigDiffPES}
\end{figure*}

Fig.~\ref{fig:FigDiffPES} compares two-dimensional cuts through the PESs of both molecules for a sGDML model trained on CCSD(T) data with the predictions of different MM-FFs (energy ranges for the different PES are reported in the SI).
The Generalized Amber Force Field (GAFF)\cite{gaff2004} is a representative of the Bixon-Lifson type FFs introduced in Sec.~\ref{MMFFs}.
The other MM-FFs, the Merck Molecular Force Field (MMFF94)~\cite{MMFF94}, Ghemical~\cite{Ghemical} and the Universal Force Field (UFF)~\cite{UFF}, are extensions to this framework and introduce more complex functional forms in order to improve generalization and predictive accuracy.
All MM-FFs studied here have been selected due to their performance for small to medium sized organic molecules. % which are the focus of this study.

% ethanol
In the case of the ethanol molecule, a qualitative comparison reveals that GAFF and MMFF94 yield good approximations to the sGDML@CCSD(T) reference PES.
However, they miss the coupling between the hydroxyl and methyl functional group caused by the interactions of the oxygen lone pair electrons with the methyl hydrogens.
This coupling induces an asymmetric behavior in the PES of both angle coordinates, which can be observed for the sGDML model.
While the absence of this effect is to be expected in GAFF, since there is no explicit description of lone pairs, MMFF94 also fails to account for the interaction despite its more elaborate functional form. 
Ghemical and UFF, on the other hand, both manage to reproduce the rotor coupling qualitatively but completely misrepresent the remainder of the PES.

% keto-MDA
The keto-MDA molecule is a particularly interesting system with two symmetrical main degrees of freedom in the form of its aldehyde rotors.
The PES associated with these rotors displays a wide variety of features due to complex quantum-mechanical interactions~\cite{sGDMLappl1}. 
Hence, accounting for these effects is nontrivial for conventional FF approaches as well as ML techniques. 
In Fig.~\ref{fig:FigPESAllFFs} we observe, that the default version of UFF is insufficiently parametrized to describe the interaction between the two aldehyde groups in general.
GAFF reproduces the global minima qualitatively well, mainly in the form of electrostatic interactions between both rotors.
Yet, as would be expected from a Brixon-Lifson FF, it misses the two local minima generated by non-classical electronic interactions.
Interestingly, both MMFF94 and Ghemical FFs generate a PES that bears only remote resemblance to the PES predicted by sGDML@CCSD(T) or GAFF.
The reason for this behavior is that both FFs were primarily parametrized using data at the Hartree-Fock (HF) level of theory~\cite{MMFF94}. 
For MMFF94 in particular, the agreement with the reference HF PES as depicted in Fig.~\ref{fig:FigCompHF} is remarkable.
This is an important result, since it demonstrates that the more complex functional form of the MMFF94 FF could in principle be reparametrized using higher level of theory data (generated e.g. with the sGDML@CCSD(T) model) to yield high quality PES.

% ==================== F I G U R E ===================
% =====================================================

In the following, we perform a more detailed analysis of the sGDML@CCSD(T) model and GAFF as representative for MM-FFs in order to gain a better understanding of the strengths and shortcomings of both approaches.
Such an in-depth understanding is tantamount for devising potential improvements to both kinds of FFs.
GAFF is chosen as the primary MM-FF, due to its general reliability (see above), as well as its simple analytical form which facilitates the interpretation of results.
Fig.~\ref{fig:FigDiffPES} shows detailed PESs of ethanol and keto-MDA for sGDML@CCSD(T) and GAFF, together with the differences between both methods.

% Ethanol
In the case of the ethanol molecule (Fig.~\ref{fig:FigDiffPES}, first row), the deviations between both approaches appear to be only minor when comparing the PES directly.
However, the difference plot reveals the details missing in the GAFF description in the form of the coupling between the two rotors in ethanol.
% MDA
The overall differences are more obvious for keto-MDA, where only the global minima and the central peak in the PES are reproduced satisfyingly by GAFF (see Fig.~\ref{fig:FigDiffPES}, second row).
The fact that, in general, GAFF describes the reference PESs qualitatively well for these two molecules (in particular their global minimum), opens the opportunity to use it for creating a reliable sampling of their PES via molecular dynamics.
This would be an alternative to its more computationally expensive counterpart, DFT molecular dynamics, as we will explore in section~\ref{CompDetails}.

% ==================== F I G U R E ===================
\begin{figure}[ht]
	\centering
	\includegraphics[width=1.0\columnwidth]{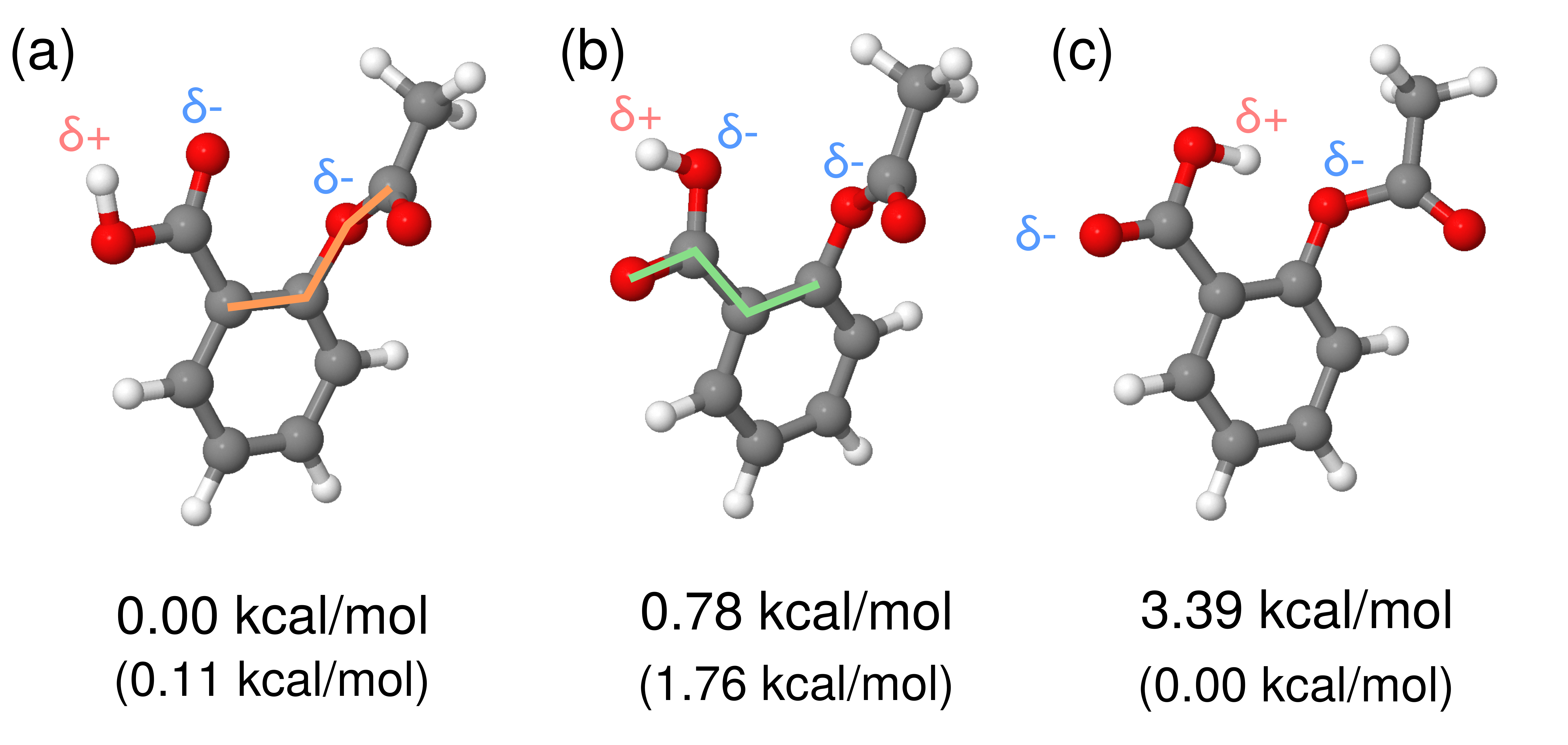}
	\caption{Three lowest minima of aspirin computed with CCSD/cc-pVDZ. Energy differences relative to the minimum in kcal\,mol$^{-1}$ are shown for the sGDML@CCSD model and GAFF (in parenthesis). }
	\label{fig:FigminAsp}
\end{figure}
% =====================================================

In addition to ethanol and keto-MDA, we also study the performance of MM-FFs for the aspirin molecule.
Here, we use GAFF to predict the ordering of the three lowest lying equilibrium configurations of aspirin identified with a sGDML@CCSD model in a previous study~\cite{sgdml}.
The associated structures, along with the relative sGDML@CCSD energy differences, are depicted in Fig.~\ref{fig:FigminAsp}.
Contrary to the cases of ethanol and keto-MDA, GAFF strongly misrepresents the ordering of the minima on the aspirin PES (Fig.~\ref{fig:FigminAsp}, parenthesis).
In aspirin the global minimum shown in Fig.~\ref{fig:FigminAsp}-a is stabilized by the $n\to\pi^*$ attractive interaction between the electron lone pair $n$ of the oxygen atom in the carboxylic acid group and the anti-bonding $\pi^*$ orbital located in the carbon atom of the ester group~\cite{n2pi_Aspirin2011}.
If we remove this interaction from the equation and just consider the electrostatic contributions such in the GAFF case, the geometric configuration Fig.~\ref{fig:FigminAsp}-a is not longer stable due to the very close proximity of the three negatively charged oxygen atoms. 
In this case, Fig.~\ref{fig:FigminAsp}-c is the most stable configuration, where two the oxygen atoms are sharing a positive hydrogen atom. In the CCSD level of theory, this structure is also a local minimum, but it is 3.39~kcal\,mol$^{-1}$ higher than the global minimum. 

\subsection{Anharmonicities and Vibrational Spectra}\label{Spectra}

\begin{figure*}[ht]
	\centering
	\includegraphics[width=0.9\textwidth]{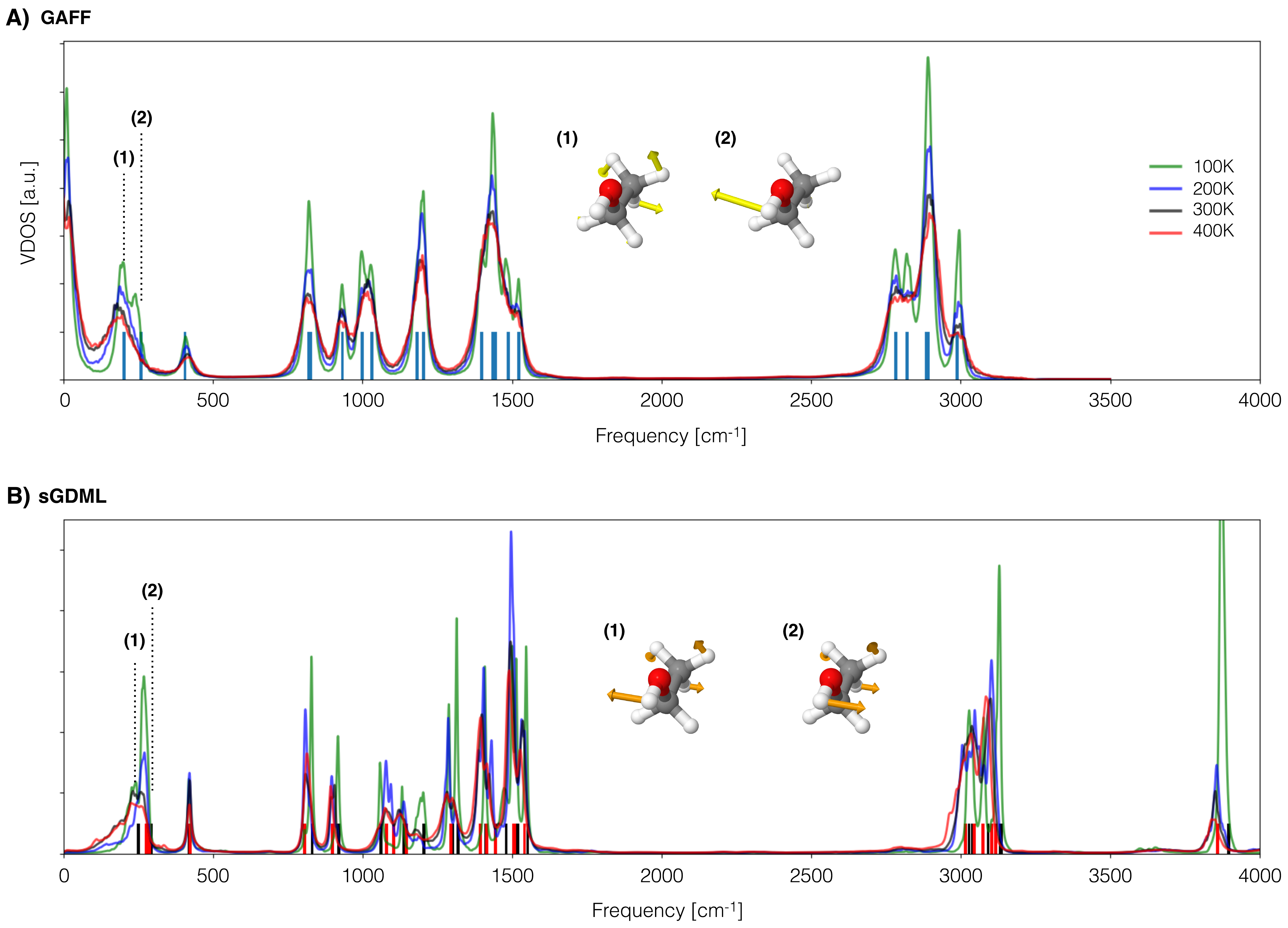}
	\caption{Comparison of the VDOSs generated with sGDML@CCSD(T) and GAFF force fields for ethanol.}
	\label{fig:FigVDOS}
\end{figure*}

To supplement the previous analysis based on a static picture, we want to briefly discuss some of the dynamical implications of using 
%To close this section, we want to shortly discuss some of the dynamical implications of using
conventional FFs in general. We performed a series of long MD trajectories of the ethanol molecule for different temperatures (100 to 400K) to compute the vibrational density of states (VDOS). 
In Fig.~\ref{fig:FigVDOS} we show the results for GAFF and for sGDML@CCSD(T).
% ==================== F I G U R E ===================
% =====================================================
%
Firstly, we can see that GAFF gives a good prediction of the VDOS in the fingerprint region (500 to 1500\,cm$^{-1}$), since in general it is parametrized to perform well in this frequency range. 
On the other hand, it poorly describes the bond stretching vibrations involving hydrogen atoms ($\sim$3000\,cm$^{-1}$). 
Moreover, if we analyze the evolution of the VDOS with temperature, we can see that there is no major change or shift in the GAFF frequencies with increasing temperature. 
This a direct consequence of the MM-FFs functional form, where bond and angle interactions (Eqs.~\ref{eq:bondharm} and \ref{eq:angharm}) are approximated as harmonic terms, neglecting higher order contributions.
However, in reality both interactions cannot be expressed purely in terms of harmonic oscillators, with bond dissociation being one of the best counterexamples.
These anharmonic contributions give rise to the temperature dependent shifts absent in the middle to high frequency regimes of the MM-FF VDOS ($>$500\,cm$^{-1}$), but modeled correctly by the ML-FF.
For low frequency vibrations, anharmonicities can still be captured by a MM-FF through the interplay of dihedral terms (Eq.~\ref{eq:dihed}) and long range interactions (Eqs.~\ref{eq:coulomb} and \ref{eq:lj}) which are only computed between atoms at least three bonds removed in classical Bixon--Lifson FFs.
This behavior can be observed in the VDOS regions of the hydroxyl and methyl rotors ($<$500\,cm$^{-1}$), which both show a temperature induced red-shift.
The dynamics obtained using the sGDML@CCSD(T), on the other hand, exhibit non trivial shifts in the frequencies and VDOS populations across the whole spectrum, demonstrating the advantage of the less restricted functional forms used in ML-FFs.

It is worth to note, that by analyzing the vibrational normal modes we can see that the two rotors are completely decoupled in the case of GAFF, while in the sGDML case the well known coupling is present. 
Based on this, we conclude that the lack of anharmonicities in GAFF will certainly affect the sampling of the PES, as will be explored in the following. 

%%%%%%%%%%%%%%%%%%%%%%%%%%%%%%%%%%%%%%%%%%%%%%%%%
%============= Database construction ============%
%%%%%%%%%%%%%%%%%%%%%%%%%%%%%%%%%%%%%%%%%%%%%%%%%
\subsection{Generating Reference Data with MM-FFs} \label{CompDetails}

For machine learning methods, the construction of suitable reference databases is a crucial step, given that any model trained or fitted to a given database can only generate predictions as accurate as the data itself. 
For example, a model trained on DFT data, can at most give results as accurate as DFT. 
Therefore, the level of theory and methodology used to generate the data has to be chosen carefully, trying to maximize the amount of data accessible at a theory level that accurately encodes the quantum-mechanical effects of interest utilizing the computational resources at hand.
% DFT data
In this regard, DFT reference data generated by running molecular dynamics simulations at a high temperature (500K) has for example been used to generate large and reliable molecular datasets, known as MD17~\cite{gdml,sgdml,sGDMLappl1}.
The software used to generate the MD17 dataset was the FHI-aims package~\cite{FHIaims2009} using generalized gradient approximation (GGA) with Perdew-Burke-Ernzerhof (PBE)~\cite{PBE1996} exchange-correlation functional combined with the Tkatchenko-Scheffler (TS) method~\cite{TS} to incorporate van der Waals interactions.

While a direct sampling approach is feasible at the DFT level, it quickly become intractable at higher levels of theory.
To circumvent this limitation, a two-step data generation approach can be used to create more accurate datasets (see e.g. Ref.~\citenum{behler2015constructing}). 
In a recent example, such a procedure was used to generate a CCSD(T) level MD17 dataset, by randomly subsampling some of the molecular dynamics trajectories in the MD17 dataset and recomputing energies and forces using CCSD(T) level of theory.
This methodology successfully generated highly accurate sGDML@CCSD(T) FFs able to replicate experimental results giving further insights about the importance of electron correlation and nuclear quantum effects~\cite{sgdml,sGDMLappl1}.
The origin of the success of the two-step process is that the PES$_{\text{DFT}}$ sampled in the MD17 trajectories contains already most of the general features in the PES$_{\text{CCSD(T)}}$, therefore the sGDML model can use the same molecular configurations as basis to reconstruct both PESs.

This naturally raises the question on the validity of using datasets generated by MD simulations with  conventional FFs as basis for a similar two-step reference database construction.
Such a procedure has long been suggested, since it would offer the opportunity to create vast amounts of molecular configurations to be used as inputs in machine learning methods.
Nevertheless, to our knowledge, no study has ever analyzed the reliability of the set of configurations generated in such a manner.
As stated above, the success of a two step procedure would depend on how similar the MM-FF free energy is to the ones generated by the reference electronic-structure method, in our case GAFF and DFT or CCSD(T), respectively.
In the previous sections we already analyzed some of the differences, advantages, disadvantages, regimes of applicability and possible problems for certain types of molecules.
%

\begin{figure*}[ht]
\centering
\includegraphics[width=0.9\textwidth]{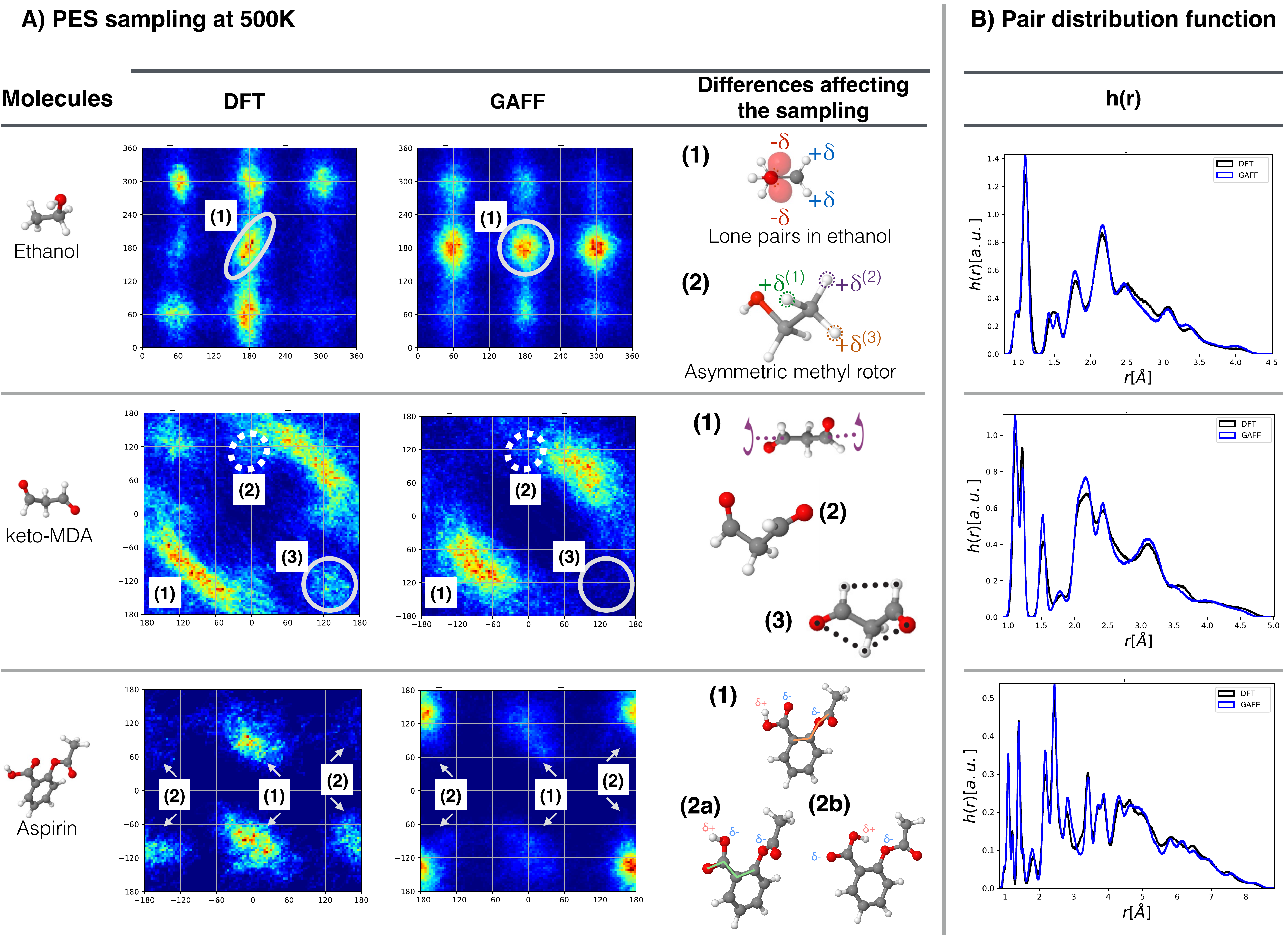}
\caption{Sampling generated by Comparison of sampling generated form DFT and GAFF-force field for ethanol, keto-MDA and aspirin. The overlap between the sampling is shown in the rightmost column as a pair distribution function of interatomic distances.}
\label{fig:FigSampling}
\end{figure*}

% TODO: transiton, remove MD17AMBER here.
With this information at hand, we now explore how these effects influence the sampling behavior of GAFF and sGDML, with a particular focus on the implications this has for using GAFF to construct a MM-FF based version of the MD17 dataset. 
In Fig.~\ref{fig:FigSampling} we present a comparison between the sampling generated by DFT and GAFF molecular dynamics for ethanol, keto-MDA and aspirin at 500K. 
From Fig.~\ref{fig:FigSampling}-A is straightforward to see that, even in this 2D representation of the data, these two methodologies generate different samplings of the configurational space. 
As mentioned before, in the case of ethanol one of the key ingredients missing in GAFF is the coupling between methyl and hydroxyl rotors and the fact that the methyl rotor is not symmetric due to its parametrization. 
Of course, this is a projection of the data in their two main degrees of freedom, therefore in Fig.~\ref{fig:FigSampling}-B we show the radial distribution function comparing the interatomic distances obtained in each case. 
Clearly, it was not expected for GAFF to generate the same results as DFT, but the sole purpose of this section is to highlight the different dynamics generated by each approximation. 
From Fig.~\ref{fig:FigSampling}-B we can see that in general GAFF performs with reliable accuracy in predicting the interatomic distances. 
The second row of Fig.~\ref{fig:FigSampling}-A is the demonstration to what was discussed earlier in section~\ref{PES} in terms of the PES of keto-MDA, where the missing local minima will influence the sampling process. 
The most radical case is aspirin, where GAFF preferably samples two of its local minima instead of the global one, as shown in Fig.~\ref{fig:FigSampling}-A bottom.

Now, from Fig.~\ref{fig:FigSampling} we have evidence that DFT and GAFF, in general, sample different parts of the configuration space.
Therefore, with this information at hand the next question to answer is whether the GAFF sampled configurations can still be used to reconstruct the DFT PES?
It should also be noted at this point, that the goal of the learning process is to generate a function able to generalize, i.e. to reconstruct, the underlying level of theory used to generate the database and not to learn or memorize the database itself.
%

 % ==================== F I G U R E ===================
% =====================================================

% ~~~~~~~~~~~~~~~~~~~~~~~~~~~~~~~~~~~~~~~~~~~~~~~~~~~~~~~~~~~~~~~~~~~~~~

\subsection{Cross Comparison between MM-FF and DFT sampling}

Creating highly accurate \textit{ab-initio} (e.g. CCSD(T)) data directly from MD simulations is infeasible for all but the smallest molecules.
Instead it is customary to run the simulations at a lower and hence more affordable level of theory such as DFT (e.g PBE) combined with a small basis set.
A fundamental hypothesis is, that the PES generated by DFT methods is close enough to a higher level of theory PES such as CCSD(T).
Therefore, by sampling the PES using DFT based molecular dynamics simulations, one would expect the generated molecular configurations to be a good approximation to sampling the corresponding CCSD(T) PES directly.
This in turn makes it possible to accurately reconstruct the CCSD(T) PES from MD@DFT trajectories. 
This has been empirically shown to be a valid assumption~\cite{sgdml,sGDMLsoftware2019,sGDMLappl1,NQE_vdW_2020}, since DFT functionals incorporate a considerable amount of exchange and correlation to describe the quantum system.

\begin{figure}[ht]
\centering
\includegraphics[width=1.0\columnwidth]{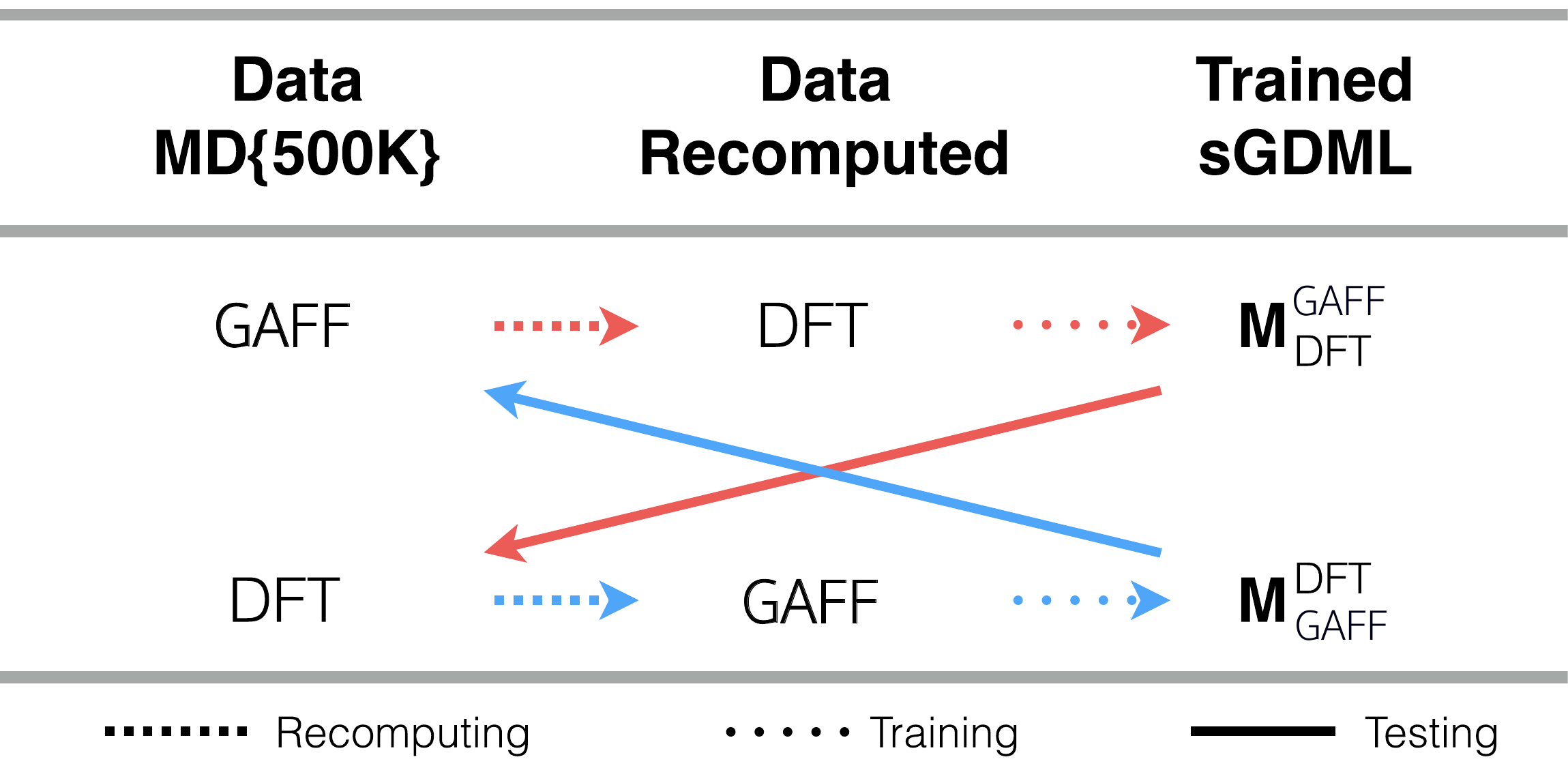}
\caption{Scheme describing the generation of datasets and models used for the cross comparison experiments.}
\label{fig:FigResampling}
\end{figure}

It is tempting to use the same approach to generate data from MD@ML-FFs to reconstruct DFT PESs, but from Figs.~\ref{fig:FigPESAllFFs} and~\ref{fig:FigSampling}  we see that the GAFF and PBE+TS PESs exhibit fundamental differences.
Hence, running MD simulations on each surface will generate distinct populations of configuration space~\cite{sgdml}. 
Consequently, there is a high chance that the PES reconstructed by the sGDML model based on MD@FFs data will generate good training errors, but will not be able to generalize properly. 
In order to verify this assertion, we randomly sampled 2000 geometries (1000 for training and 1000 for testing) from the GAFF simulations used in Fig.~\ref{fig:FigSampling}, which were recomputed using DFT (PBE+TS, see Supplementary information for more details).
This data, which we refer to as the MD17$^{\text{GAFF}}$ dataset, was then used to train a set of sGDML models (see Fig.~\ref{fig:FigResampling}).
%A common idea is to use regular FFs to generate big amounts of data (geometries) to then sub-sample and recompute with an \textit{ab-initio} method to train ML models. 
The results of this experiment are reported in Table~\ref{tab:ccsd_results}.

\begin{table}[ht]
\caption{Prediction accuracy for interatomic forces of the GAFF sampled datasets at 500K and recomputed using PBE+TS. Force errors are reported in kcal $\text{mol}^{-1} \text{\AA}^{-1}$. }
\label{tab:ccsd_results}
\setlength\extrarowheight{1pt}
\begin{ruledtabular}
\begin{tabular}{lllrr}
\multirow{2}{*}{Dataset} & \multicolumn{2}{c}{Sampling} & \multicolumn{2}{c}{Force Prediction}\\
\cline{2-3} \cline{4-5}
                          & Training & Validation        & MAE  & RMSE \\
\hline
\multirow{3}{*}{keto-MDA} & GAFF     & GAFF              & 0.29 & 0.44 \\
                          & DFT      & DFT               & 0.41 & 0.62 \\
                          & GAFF     & DFT               & 0.54 & 0.82 \\
\hline
\multirow{3}{*}{ethanol}  & GAFF     & GAFF              & 0.27 & 0.42 \\
                          & DFT      & DFT               & 0.33 & 0.49 \\
                          & GAFF     & DFT               & 0.45 & 0.71 \\
\end{tabular}
\end{ruledtabular}
\end{table}

Based on these results, we find that for the considered molecules the sGDML models trained and validated on GAFF data (sGDML@MD17$^{\text{GAFF}}$) perform slightly better that the sGDML models trained and validated on DFT data (sGDML@MD17$^{\text{DFT}}$). 
Nevertheless, when validating the sGDML@MD17$^{\text{GAFF}}$ model on MD@DFT sampled data, the errors increase almost by a factor of two.
We would like to highlight that even when validating the sGDML@MD17$^{\text{GAFF}}$ model on unseen data from another distribution it manages to generate good force MAEs below 0.9 kcal mol$^{-1}$ \AA$^{-1}$ which is a remarkable achievement.
This behavior hints at a potential use in conjunction with adaptive sampling approaches\cite{Gastegger2017}, where an initial model trained on MM-FF sampled data could serve as a basis for further iterative refinement.

The analysis presented here, gives a better perspective on the use of conventional MM-FFs to generate representative databases to higher levels of theory for ML reconstruction of the molecular PES.

% TODO: Caption
% ==================== F I G U R E ===================
% =====================================================

\subsection{Augmenting MM-FFs with ML}\label{Reparametrization}

Up to this point, we have focused on GAFF, a Bixon and Lifson MM-FF with predetermined functional form and parametrization scheme.
However, as can be seen in Fig.~\ref{fig:FigDiffPES}, extending the functional form or changing the set of parameters can lead to significant differences in the behavior of force fields.
This hints at potential applications where the synergy between MM-FFs and ML approaches can be exploited in order to improve upon conventional FFs.
Here, we study aspects of two possible MM/ML-FF setups, where a) ML methods might be used to generate additional data for parametrizing a MM-FF at a higher level of theory or b) augment the functional form of a force field, incorporating e.g. interactions due to quantum effects.
%In the following, we study different aspects of these potential interactions between ML and MM-FFs.

In a first series of experiments, we explore to what extent MM-FFs can profit from additional data as well as changes to their functional makeup.
%And how the functional form of a FF influences its ability to utilize this data.
%In a final experiment, we study the data saturation and influence of the exact functional form on the performance of classical FFs.
To this end, we construct GAFF type force-fields for ethanol and aspirin, based on the electronic-structure energies and forces of 110, 1100 and 11\,000 reference structures taken from the MD17 database.
As a baseline MM-FF, we introduce a PyTorch\cite{NEURIPS2019_9015} implementation of the GAFF.
This allows us to reoptimize all GAFF parameters using molecular energies and forces, as would be done in a typical ML-FF scenario.
Moreover, in order to study the effect of changing the functional form, we introduce two extensions to this basic GAFF.
In the first variant, all harmonic bond potentials are replaced by Morse potentials and the number of terms in the dihedral Fourier series is doubled (see Sec.~\ref{MMFFs}). 
The second variant introduces additional flexibility, by modeling all bonded interactions via feed-forward neural networks (NNs) instead of the conventional physical approximations (see SI for more details).
These models will be refered to as GAFF-Morse and GAFF-NN, respectively.
The test set errors achieved for the differently sized training sets, molecules and GAFF variants are reported in Table~\ref{tab:mlff}.
Shown are the energy and force RMSEs obtained as the average over three independent runs for each model.
\begin{table*}[ht]	
\caption{Classical and modified GAFFs trained on MD17 reference data for ethanol and aspirin. GAFF denotes the unmodified GAFF force field, while GAFF-Morse and GAFF-NN corresponds to the variants using Morse potentials and NNs. Energy and force RMSEs on the test set are given in kcal $\text{mol}^{-1}$ and kcal $\text{mol}^{-1} \text{\AA}^{-1}$, respectively. All errors are obtained as the average over three independent training runs. Where available, we have added the sGDML results.}
\label{tab:mlff}
\begin{ruledtabular}
\begin{tabular}{llrrrrrrrrr}
&            & \multicolumn{4}{c}{Energy} && \multicolumn{4}{c}{Forces}\\
\cline{3-6} \cline{7-11}
Molecule & Datapoints & sGDML & GAFF & GAFF-Morse & GAFF-NN && sGDML & GAFF & GAFF-Morse & GAFF-NN\\
\hline
\multirow{3}{10pt}{Ethanol} &   110 & --    & 1.27 & 2.40 & 1.37 && --    & 6.53 & 15.38 & 5.00 \\
							&  1100 & 0.07  & 1.21 & 2.51 & 1.16 && 0.34 & 6.37 & 14.59 & 4.78 \\
							& 11000 & --    & 1.20 & 1.15 & 1.11 && --    & 6.38 &  4.80 & 4.66 \\
\hline
\multirow{3}{10pt}{Aspirin} &   110 & --    & 2.18 & 2.94 & 2.28 && --    & 7.91 & 10.29 & 6.72 \\
							&  1100 & 0.19 & 2.20 & 2.37 & 2.19 && 0.68 & 7.46 &  7.77 & 6.15 \\
							& 11000 & --    & 2.05 & 2.10 & 1.98 && --    & 7.35 &  6.28 & 5.95 \\
\end{tabular}
\end{ruledtabular}
\end{table*}

% GAFF refit
In general, significantly higher errors are obtained for the GAFF variants compared to pure ML-FFs, such as sGDML.
This is caused by the less flexible, physically motivated form of MM-FFs, compared to their ML counterpart.
However, the potential advantage of constructing force fields in such a manner becomes apparent when looking at the saturation of the error with respect to the amount of reference data used during training.
Even with only 110 data points, MM-FFs exhibit good generalization performance, which changes little upon the inclusion of more data.
This is especially remarkable in the case of flexible molecules such as aspirin.
These observations indicate, that the refitted GAFF models for ethanol and aspirin studied here are already saturated with respect to the training data and little performance can be gained upon providing additional data without fundamentally changing the functional form of classical FFs.
% GAFF + Morse
Moving on to GAFF-Morse, we find the above conclusions confirmed.
The increase in functional flexibility allows the model to converge to slightly better RMSEs compared to the unmodified versions, as well as utilize additional data in a more effective manner.
It is also interesting to note, that forces appear to profit from the extended functional form to a larger extent than the energies.
The reason for this effect is the presence of anharmonic contributions to the bond energies, also observed in the VDOSs studied in Sec.~\ref{Spectra}.
% GAFF + NNs
As would be expected from its more flexible NN based structure, GAFF-NN continues the trends observed for data saturation and predictive accuracy.
Energies and force predictions in particular appear to benefit greatly from the more flexible functional form provided by the NNs.
Interestingly, even in the case of using only 110 data points, GAFF-NN outperforms the other variants with respect to force predictions.
This suggests, that MM methods can profit from ML based terms even in low data regimes.
%This suggests, that ML methods can profit from integrating them into the typical MM-FF structure in general, potentially improving their generalization capabilities.

\begin{figure*}[ht]
	\centering
	\includegraphics[width=0.9\textwidth]{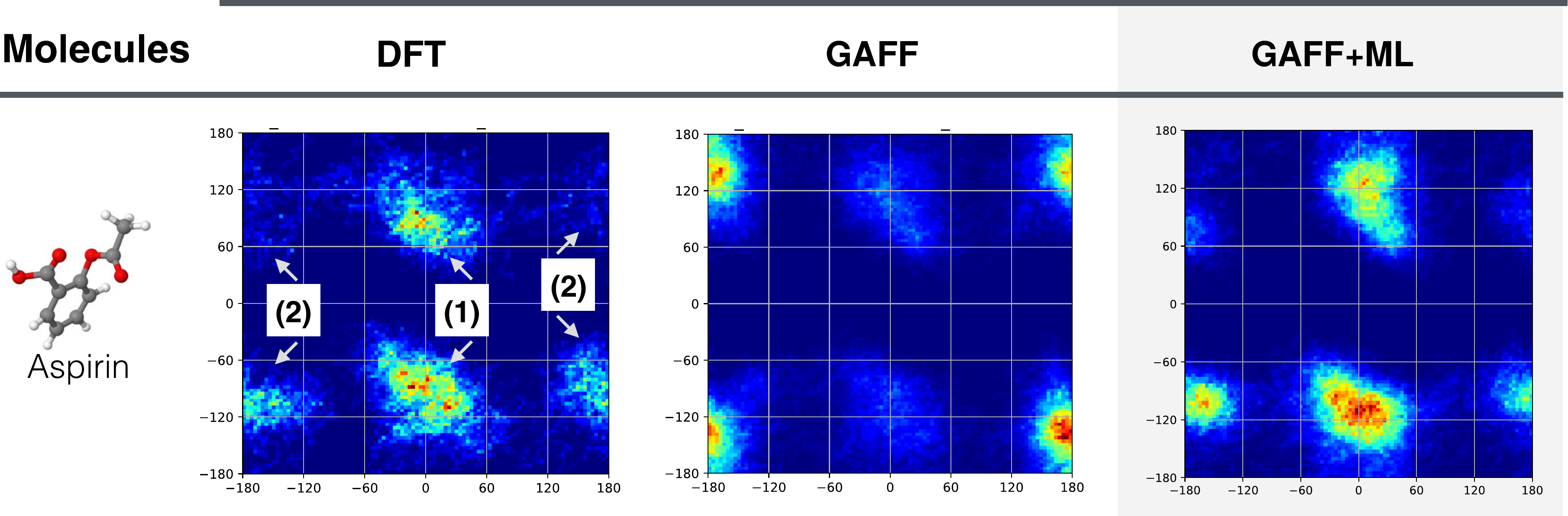}
	\caption{Comparison of sampling of the three aspirin minima by a sGDML@DFT model, standard GAFF and a GAFF model corrected for the ${n\to\pi*}$ interaction term.}
	\label{fig:FigCorrection}
\end{figure*}

In a final experiment, we perform a proof of principle study on how ML approaches can be used to formulate extensions to MM-FFs, compensating e.g. for missing interaction terms.
To this end, we target the ${n\to\pi*}$ interaction in the aspirin molecule as a test application.
The inability to account for this quantum-mechanical effect is the reason for GAFF yielding the wrong ordering of minima (Fig~\ref{fig:FigminAsp}) and exhibiting qualitatively different sampling (see Fig.~\ref{fig:FigSampling}).
We introduce a sGDML based extension parametrized to correctly capture the orbital-orbital interaction (see SI for details).
A GAFF model augmented with this ML term, $\mathbf{\hat{f}_{F}}^{(n\to\pi*)}$, is then used to repeat the sampling experiments performed in Sec.~\ref{CompDetails}.
Fig.~\ref{fig:FigCorrection} shows the different sampling behavior of sGDML@CCSD, pure GAFF and the extended GAFF model (GAFF-ML).

% ==================== F I G U R E ===================
% =====================================================

As can be seen, even a relative simple augmentation is enough to recover the sampling behavior of the reference method to a large extent.
The global minimum dominated by the ${n\to\pi*}$ interaction (1) is now visited preferably by the GAFF+ML model, which agrees with results from sGDML@CCSD~\cite{NQE_vdW_2020}.
The relative sampling frequency of visiting the different minima is reproduced as well.
This result is particularly impressive, considering the restrictive functional form of the basic MM-FF.
Even more promising is the procedure by which this term was introduced, exemplifying a potential synergistic approach to developing force fields, where missing terms can be identified via ML-FF comparisons, condensed into a ML model and then deployed with MM-FFs in a highly systematic manner.

\section{Conclusions and Outlook}\label{Concl}

Machine learning based force fields (ML-FFs) show excellent accuracy in modeling a wide range of chemical phenomena.
Using specialized architectures, such as sGDML studied here, extremely reliable surrogate models of high-level electronic-structure methods can be constructed.
These models are able to capture a wide range of the underlying quantum many-body effects, while at the same time only incurring a fraction of the computational cost of the original method.
However, due the involved functional structure used, the computational speed of ML-FFs still pales in comparison to classical molecular mechanics force fields (MM-FFs).
MM-FFs employ a set of highly efficient approximations, making them capable of simulating systems containing many thousands of atoms.
However, this efficiency comes at the cost of general accuracy, rendering MM-FFs less reliable or even incapable of modeling complex chemical phenomena.
The focus of the present study was to identify differences and synergies between MM-FFs and ML-FFs and how the resulting knowledge can be used to combine both approaches.
%The focus of this present study is how the synergies between both approaches -- MM-FFs and ML-FFs -- can be exploited in order to improve both approaches and even develop new models, further pushing back the boundaries of what can be simulated with FFs.

% Conclusions?
An important aspect of such a synergistic approach is the analysis of models and identifications of their shortcomings.
Here, we could show that accurate ML-FFs offer access to high-quality PES at the CCSD(T) level of theory, which can be used to probe the performance of MM-FFs in a very efficient manner.
In this way, insights can be gained into situations where models perform reliably and cases where they fail completely.
Moreover, ML-FFs can not only be used for a general analysis, but even allow to pinpoint physical phenomena systematically missing from the description of MM-FFs, thus guiding the development of new, improved approaches.
Another aspect explored in this study is the use of MM-FFs for generating reference data sets to be used in constructing ML-FFs.
Due to their high computational efficiency, MM-FFs constitute a powerful tool for sampling complex potential-energy surfaces (PESs).
It was found, that MM-FFs can be used to directly sample reference configurations for ML models parametrized at a high electronic-structure level, provided a sufficiently high overlap between the PES of both methods is present.
However, even in situations where this assumption does not hold, MM-FFs show acceptable performance which makes them highly promising in the early stages of constructing ML-FFs as they can e.g. provide an initial set of reference configurations which can then be further refined with adaptive sampling \cite{Gastegger2017}.
ML approaches can also be used to directly augment MM-FFs.
Here, we have investigated how changes to the functional form influence their capacity to leverage data in varying amounts of reference data.
Increasing the flexibility of MM-FFs by modeling all bonded interactions via neural networks, yielded a model which exhibits improved accuracy over conventional schemes already in small data regimes and consistently improves upon addition of more data.
This suggests, that MM-FFs can profit greatly from functional features of modern ML algorithms.
These can either be used in a targeted fashion as corrections to include additional interactions not captured by the basic force field or tightly integrated into a MM-FFs structure.
The more flexible functional forms allow MM-FF models to capture a wider range of phenomena and leverage additional information without compromising the basic design paradigm of MM-FFs.
In addition, the experiments in this study also demonstrated how ML components can in turn benefit from being embedded in a MM-FF framework.
The principled physical structure and decomposition of interactions has the potential to further improve data efficiency and generalization behavior of ML-FFs.

All these findings point towards the general potential of integrating force fields and ML approaches.
Efficient hybrid approaches offering high performance could be constructed, where ML-FF terms restricted to local regions of a system could be introduced to MM-FFs, accounting for physical phenomena not fully described by the purely classical interaction, e.g. orbital based effects.
This could for example be achieved by restricting the molecular descriptor in sGDML to a set of atoms, similar to how bonded interactions are treated in Bixon-Lifson force fields.
Another important phenomenon, where terms derived from ML-FFs could prove highly advantageous is the treatment of long range van der Waals interactions.
These are typically accounted for in an additive pair-wise manner.
However, there is mounting evidence that many-body effects play an important role for these interactions, which might even be crucial in describing the macroscopic structure of soft matter \cite{hermann2017first}.
Other potential applications where both -- MM-FF and ML-FF -- could be combined to great effect is the construction of ML-FFs for extended systems, such as proteins or solutions.
There, generating high-level reference data via sampling driven purely by electronic-structure theory is infeasible.
MM-FFs on the other hand, could provide efficient access to large parts of the configuration space.
Missing regions could then be sampled iteratively.

In conclusion, classical MM-FFs and modern ML-FFs both exhibit particular strengths complementing each other.
Here, we have investigated various facets of how both approaches can be integrated with each other, taking a first step towards systematically combining their strengths.

\section*{Data Availability Statement}
The training data  and code base used in this study are available from the authors upon reasonable request.

\begin{acknowledgments}
This project has received funding from the European Unions Horizon 2020 research and innovation program under the Marie Sk\l{}odowska-Curie grant agreement No. 792572 (M.G.).
KRM acknowledges financial support by the German Ministry for Education and Research (BMBF) for the Berlin Center for Machine Learning (01IS18037A) and under the Grants 01IS14013A-E, 01GQ1115 and 01GQ0850; Deutsche Forschungsgemeinschaft (DFG) under Grant Math+, EXC 2046/1, Project ID 390685689 and by the Technology Promotion (IITP) grant funded by the Korea government (No. 2017-0-00451, No. 2017-0-01779). Correspondence to KRM and AT.
\end{acknowledgments}

\bibliography{references/references}% Produces the bibliography via BibTeX.

\end{document}